# Identification of Protein Coding Regions in Genomic DNA Using Unsupervised FMACA Based Pattern Classifier

P. Kiran Sree[†]  Dr. Inampudi Ramesh Babu[††]

[†] Associate Professor, Department of Computer Science,
S.R.K Institute of Technology, Enikepadu, Vijayawada.INDIA
[††] Senior IEEE member & Professor, Department of Computer Science,
Acharya Nagarjuna University, Guntur, India.

**Abstract**

Genes carry the instructions for making proteins that are found in a cell as a specific sequence of nucleotides that are found in DNA molecules. But, the regions of these genes that code for proteins may occupy only a small region of the sequence. Identifying the coding regions play a vital role in understanding these genes. In this paper we propose a unsupervised Fuzzy Multiple Attractor Cellular Automata (FMCA) based pattern classifier to identify the coding region of a DNA sequence. We propose a distinct K-Means algorithm for designing FMACA classifier which is simple, efficient and produces more accurate classifier than that has previously been obtained for a range of different sequence lengths. Experimental results confirm the scalability of the proposed Unsupervised FCA based classifier to handle large volume of datasets irrespective of the number of classes, tuples and attributes. Good classification accuracy has been established.

## 1. Introduction

Many of the challenges in biology are now challenges in computing. Bioinformatics, the application of computational techniques to analyze the information associated with bimolecules on a large scale, has now firmly established itself as a discipline in molecular biology. Bioinformatics is a management information system for molecular biology. Bioinformatics encompasses everything from data storage and retrieval to the identification and presentation of features within data, such as finding genes within DNA sequence, finding similarities between sequences, structural predictions. Analyzing the coding regions is not the scope of the project.

For better understanding of the specified objectives, we presented CA, FCA fundamentals in Section II, Section III covers unsupervised learning as well as distinct K-Means algorithm, Section IV presents the design of FMACA based pattern classifier [3], [7] as well as rule formation and chromosome representation. In Section V, we address the problem of protein coding region identification [11], [12] in DNA sequences. In order to validate the design of proposed model, experimental results are also reported in this section.

## 2. Cellular Automata (CA) and Fuzzy Cellular Automata (FCA)

A CA [4], [5], [6], consists of a number of cells organized in the form of a lattice. It evolves in discrete space and time. The next state of a cell depends on its own state and the states of its neighboring cells. In a 3-neighborhood dependency, the next state $q_i(t+1)$ of a cell is assumed to be dependent only on itself and on its two neighbors (left and right), and is denoted as

$$q_i(t+1) = f(q_{i-1}(t), q_i(t), q_{i+1}(t)) \quad \text{-----E(1)}$$

where $q_i(t)$ represents the state of the $i^{th}$ cell at $t^{th}$ instant of time, $f$ is the next state function and referred to as the rule of the automata. The decimal equivalent of the next state function, as introduced by Wolfram, is the rule number of the CA cell. In a 2-state 3-neighborhood CA, there are total 256 distinct next state functions.

### 2.1 FCA Fundamentals

FCA [2], [6] is a linear array of cells which evolves in time. Each cell of the array assumes a state $q_i$, a rational value in the interval [0, 1] (fuzzy states) and changes its state according to a local evolution function on its own state and the states of its two neighbors. The degree to which a cell is in fuzzy states 1 and 0 can be calculated with the membership functions. This gives more accuracy in finding the coding regions. In a FCA, the conventional Boolean functions are AND , OR, NOT.

306     IJCSNS International Journal of Computer Science and Network Security, VOL.8 No.1, January 2008

## 2.2 Dependency Matrix for FCA

Rules defined in equations 1, 2 should be represented as a local transition function of FCA cell. That rules are converted into matrix form for easier representation of chromosomes [16].

| Non-complemented Rules | | Complemented Rules | |
|---|---|---|---|
| Rule | Next State | Rule | Next State |
| 0 | 0 | 255 | 1 |
| 170 | $q_{i+1}$ | 85 | $\bar{q}_{i+1}$ |
| 204 | $q_i$ | 51 | $\bar{q}_i$ |
| 238 | $q_i + q_{i+1}$ | 17 | $\overline{q_i + q_{i+1}}$ |
| 240 | $q_{i-1}$ | 15 | $\bar{q}_{i-1}$ |
| 250 | $q_{i-1} + q_{i+1}$ | 5 | $\overline{q_{i-1} + q_{i+1}}$ |
| 252 | $q_{i-1} + q_i$ | 3 | $\overline{q_{i-1} + q_i}$ |
| 254 | $q_{i-1} + q_i + q_{i+1}$ | 1 | $\overline{q_{i-1} + q_i + q_{i+1}}$ |

Table 1: FA Rules

**Example 1**: A 4-cell null boundary hybrid FCA with the following rule
< 238, 254, 238, 252 > (that is, < ($qi+qi+1$), ($qi−1+qi+qi+1$), ($qi + qi+1$), ($qi−1 + qi$) >) applied from left to right, may be characterized by the following dependency matrix

While moving from one state to other, the dependency matrix indicates on which neighboring cells the state should depend. So cell 254 depends on its state, left neighbor, and right neighbor fig (1). Now we represented the transition function in the form of matrix. In the case of complement FMACA we use another vector for representation of chromosome.

$$T = \begin{bmatrix} 1 & 1 & 0 & 0 \\ 1 & 1 & 1 & 0 \\ 0 & 0 & 1 & 1 \\ 0 & 0 & 1 & 1 \end{bmatrix}$$

Fig1: Matrix Representation

## 2.3 Transition from one state to other

Once we formulated the transition function, we can move form one state to other. For the example 1 if initial state is P (0) = (0.80, 0.20, 0.20, 0.00) then the next states will be

P (1) = (1.00 1.00, 0.20, 0.20),
P (2) = (1.00 1.00, 0.40, 0.40),
P (3) = (1.00 1.00, 0.80, 0.80),
P (4) = (1.00 1.00, 1.00, 1.00).

## 3. Unsupervised Learning

Unsupervised learning studies how systems can learn to represent particular input patterns in a way that reflects the statistical structure of the overall collection of input patterns. By contrast with supervised learning or reinforcement learning, there are no explicit target outputs or environmental evaluations associated with each input; rather the unsupervised learner brings to bear prior biases as to what aspects of the structure of the input should be captured in the output.

### 3.1 Machine learning, statistics, and information theory

Almost all work in unsupervised learning can be viewed in terms of learning a probabilistic model of the data. Even when the machine is given no supervision or reward, it may make sense for the machine to estimate a model that represents the probability distribution for a new input $x_t$ given previous inputs $x_1, \ldots, x_{t-1}$ (consider the obviously useful examples of stock prices, or the weather).

That is, the learner models $P(x_t|x_1, \ldots, x_{t-1})$. In simpler cases where the order in which the inputs arrive is irrelevant or unknown, the machine can build a model of the data which assumes that the data points $x_1, x_2, \ldots$ are independently and identically drawn from some distribution $P(x)$[2].

### 3.2 A Distinct K-Means Algorithm

The k-means algorithm with the distinct difference allows for different number of clusters, while the k-means assumes that the number of clusters is known a priori. The objective of the k-means algorithm is to minimize the within cluster variability. The objective function (which is to be minimized) is the sums of squares distances of each DNA sequence and its assigned cluster center.

$$SS_{distances} = \sum_{\forall x} [x - C(x)]^2 \quad \text{-----E(2)}$$

where C(x) is the mean of the cluster that DNA position x is assigned to Minimizing the $SS_{distances}$ is equivalent to minimizing the Mean Squared Error (MSE). The MSE is a measure of the within cluster variability.

$$MSE = \frac{\sum_i [x - C(x)]^2}{(N-c)b} = \frac{SS_{Means}}{(N-c)b} \quad \text{-----E(3)}$$

where N is the number of DNA distance centers, c indicates the number of clusters, and b is the number of spectral bands. K-means is very sensitive to initial starting values. For two classifications with different initial values and resulting different classification one could choose the classification with the smallest MSE (since this is the objective function to be minimized). However, as we show later, for two different initial values the differences in respects to the MSE are often very small while the classifications are very different. Visually it is often not clear that the classification with the smaller MSE is truly the better classification.

## 4. FMACA Based Tree-Structured Classifier

Like decision tree classifiers, FMACA based tree structured classifier uses the distinct k-means algorithm recursively partitions the training set to get nodes (attractors of a FMACA) belonging to a single class. Each node (attractor basin) of the tree is either a leaf indicating a class; or a decision (intermediate) node which specifies a test on a single FMACA, according to equations 1,2.

Suppose, we want to design a FMACA based pattern classifier to classify a training set $S = \{S1, S2, \cdot, SK\}$ into K classes. First, a FMACA with k-attractor basins is generated. The training set S is then distributed into k attractor basins (nodes). Let, S' be the set of elements in an attractor basin. If S' belongs to only one class, then label that attractor basin for that class. Otherwise, this process is repeated recursively for each attractor basin (node) until all the examples in each attractor basin belong to one class. Tree construction is reported in [7]. The above discussions have been formalized in the following algorithm. We are using genetic algorithm classify the training set.

*Algorithm 1:* **FMACA Tree Building (using distinct K means algorithms)**

Input  :   Training set $S = \{S1, S2, \cdot\cdot, SK\}$
Output:   FMACA Tree.

**Partition**(S, K)

Step 1: Generate a FMACA with k number of attractor basins.
Step 2: Distribute S into k attractor basins (nodes).
Step 3: Evaluate the distribution of examples in each attractor basin (node).
Step 4: If all the examples (S') of an attractor basin (node) belong to only one class, then label the attractor basin (leaf node) for that class.
Step 5: If examples (S') of an attractor basin belong to K' number of classes, then **Partition** (S', K').
Step 6: Stop.

## 5. Identification of Protein Coding Region in DNA Sequence

In this section we concentrate on application of FMACA to protein coding region identification. The idea of new method is to use the existing work of FMACA based tree structure classifier. Lot of research has been done for finding protein statistically. By using the standard codon frequencies, [13] we can identify whether the sequence contain protein coding regions or not.

**Example 3:**

Consider the sequence AGGACC
Since Codons will be in the form of triplets we split the input into three base sequences

So $P(S) = F$ (AGG) $\cdot F$ (ACC) $= 0.22 * 0.38 = 0.0836$ using tables from, [11], [12].
In general, Let $F0(c)$ be the frequency of codon $c$ in a non-coding sequence.
$P0 (C) = F0 (c1) F0 (c2)...F0 (cm)$
Assuming the random model of non-coding DNA, $F0(c) = 1/64 = 0.0156$ for all codons, $P0 (S) = 0.0156 \cdot 0.0156 = 0.000244$. The log-likelihood (LP) ratio for S is $LP(S) = log (0.000836/0.000244) = log (3.43) = 0.53$. If $LP(S) > 0$, **S is coding.**

Like wise we can use Bayesian classifier to calculate the probability of finding the protein coding regions with accuracy up to 49. With our approach the average accuracy achieved is 75%.

### 5.1 Data and Method

The data used for this study are the human DNA data collected by Fickett and Tung. All the sequences are taken from GenBank in May 1992. Fickett and Tung have provided the 21 different coding measures that they surveyed and compared.

The benchmark human data include three different datasets. For the first dataset, non-overlapping human DNA sequences of length 54 have been extracted from all human sequences, with shorter pieces at the ends discarded.

Every sequence is labeled according to whether it is entirely coding, entirely non-coding, or mixed, and the mixed sequences (i.e., overlapping the exon-intron boundaries) are discarded.

The dataset also includes the reverse complement of every sequence. This means that one-half of the data is guaranteed to be from the non-sense strand of the DNA.

In the next section we will give the experimental results for finding this coding region for all sequence lengths. It was compared with Un Supervised FMACA and the accuracy reported was 2.2% more than that of standard ways of finding protein coding region.

## 6. Experimental Results

The below tables shows the predictive accuracy of different algorithms on both coding and non-coding DNA sequences.

In this section we present the results on using FMACA for Fickett and Tung's dataset. Values are given for the percentage accuracy on test set coding sequences and the percentage accuracy on test set non coding sequences

**Table 2: Predictive Accuracy for length 108 human DNA Sequence**

| Algorithm | Coding | Non Coding |
|---|---|---|
| Dicodon Usage | 61% | 57% |
| Bayesian | 51% | 46% |
| CA | 78% | 72% |
| N.S.FMACA | 79% | 72.5% |

**Table 3: Predictive Accuracy for length 108 human DNA sequence**

| Algorithm | Coding | Non Coding |
|---|---|---|
| Dicodon Usage | 58% | 50% |
| Bayesian | 45% | 36% |
| CA | 74% | 69% |
| N.S.FMACA | 75% | 69% |

**Table 4: Predictive Accuracy for length 108 human DNA sequence**

| Algorithm | Coding | Non Coding |
|---|---|---|
| Dicodon Usage | 65% | 54% |
| Bayesian | 50% | 44% |
| CA | 71% | 70% |
| N.S.FMACA | 71% | 71% |

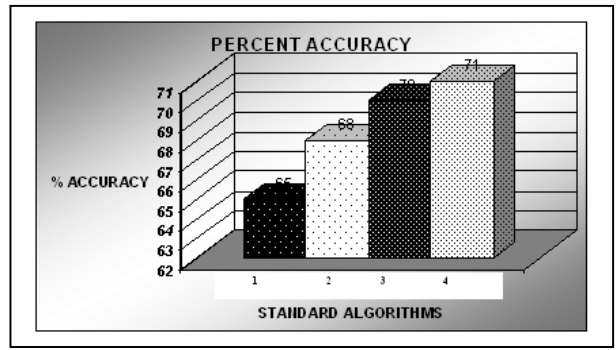

Graph 1 : Percentage Accuracy for 252 Length DNA sequence

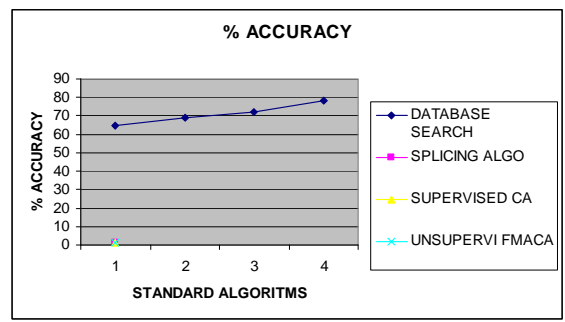

Graph 2 : Percentage Accuracy for 128 Length DNA sequence

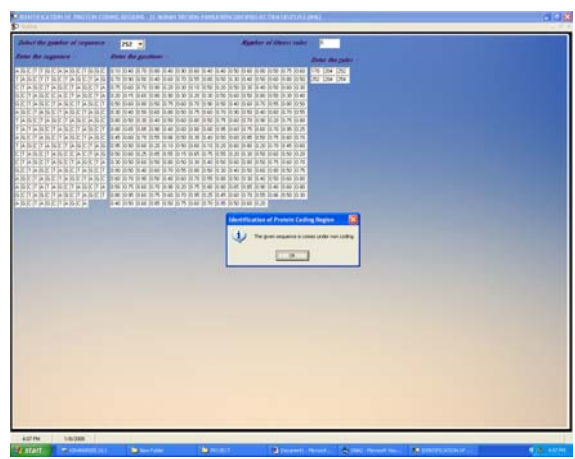

Fig: 2 UN Supervised FMACA Classifier Interface

The graphs1,2(1.Database Search ,2. Splicing Algorithm, 3. Supervised CA, 4. Un Supervised FMACA) shows Un Supervised FMACA is comparable with other three.It shows that U.S FMACA can be used to identify protein coding regions among all DNA sequence lengths. The accuracy reported also comparable with the others.

Un Supervised FMACA overcomes all the disadvantages of previous standard algorithms like fixing the position of the gene and static order of the DNA sequence. The average accuracy reported is 77%.

Fig 2 shows the Un Supervised FMACA Tool Interface, where we can find the class of a given DNA sequence.

## 7. Conclusion

This paper presents the application of MAFCA based un supervised pattern classifier to solve the problem of protein coding region identification in DNA sequences. Aside from developing a good classifier for this particular problem, the proposed model may be very much useful to solve many other bioinformatics problems like protein structure prediction, RNA structure prediction, promoter region identification, etc. .

**Acknowledgement**

I am grateful to Dr. P. Venkata Narasaiah Director, S.R.K Institute of Engineering and Technology, for his consistent support at every stage of my research work. I specially thank our President Sri B.S Apparao & Secretary Sri B.S. Sri Krishna for providing necessary infrastructure. I am indebted to my colleagues for providing wonderful atmosphere to work with.

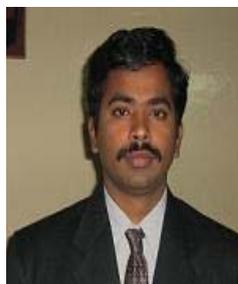

**P.KIRAN SREE** received his **B.Tech** in Computer Science & Engineering, from J.N.T.U and **M.E** in Computer Science & Engineering from Anna University. He has published many technical research papers in several international and national Journals. His areas of interests include Parallel Algorithms, Artificial Intelligence, Compile Design, Computer Networks and Cellular Automata. He also wrote books on Analysis of Algorithms, Theory of Computation and Artificial Intelligence. He is now associated with S.R.K Institute of Technology, Vijayawada.

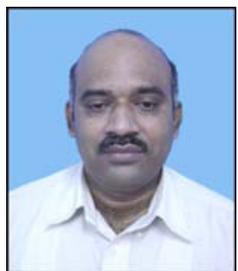

**Inampudi Ramesh Babu** received his **Ph.D** in Computer Science from Acharya Nagarjuna University, **M.E** in Computer Engineering from Andhra University, **B.E** in Electronics & Communication Engg from University of Mysore. He has published many research papers; both in international and national Journals. He is now working as Professor in Department of Computer Science at Acharya Nagarjuna University.